\documentclass[onecolumn]{revtex4-2}
\usepackage[
	bookmarks  = false,
	colorlinks = true,
	linkcolor  = blue,
	urlcolor   = blue,
	citecolor  = blue]{hyperref}
\usepackage{amsmath,amssymb,amsthm}
\usepackage{mathtools,braket}
\usepackage{graphicx,bm,bbm}
\usepackage[T1]{fontenc}
\usepackage{microtype}
\usepackage{orcidlink}
\newcommand{\appsection}[1]{\section{\MakeUppercase{#1}}}
\theoremstyle{remark}
\newtheorem{theorem}{Theorem}
\newtheorem{lemma}{Lemma}
\newtheorem{corollary}{Corollary}
\newtheorem{definition}{Definition}
\newtheorem{proposition}{Proposition}
\newtheorem{example}{Example}
\DeclareMathOperator{\Tr}{Tr}
\DeclareMathOperator{\PT}{PT}
\begin{document}
\title{Quantum data hiding with two-qubit separable states}
\author{Donghoon Ha\,\orcidlink{0000-0002-1698-9584}}
\affiliation{Department of Applied Mathematics and Institute of Natural Sciences, Kyung Hee University, Yongin 17104, Republic of Korea}
\author{Jeong San Kim\,\orcidlink{0000-0002-0420-8955}}
\email{freddie1@khu.ac.kr}
\affiliation{Department of Applied Mathematics and Institute of Natural Sciences, Kyung Hee University, Yongin 17104, Republic of Korea}
\begin{abstract}
We consider the discrimination of two-party quantum states and provide a quantum data-hiding scheme using two-qubit separable states. 
We first provide a bound on the optimal local discrimination of two-party quantum states, and establish a sufficient condition under which a two-party quantum state ensemble can be used to construct a data-hiding scheme. 
We illustrate this condition with examples of two-qubit state ensembles consisting of two orthogonal separable states.
As our data-hiding scheme can be implemented with separable states of the lowest possible dimension, its practical realization becomes significantly more attainable.
\end{abstract}
\maketitle

\section{Introduction}\label{sec:int}
Quantum data hiding is a quantum communication protocol in which information is encoded into multi-party quantum systems shared among multiple players\cite{divi2002,divi2003,lami2018}.
Under the restriction of \emph{local operations and classical communication} (LOCC), the encoded data cannot be recovered, even with unrestricted classical collaboration.
The information becomes accessible only through global operations beyond LOCC, necessitating genuinely nonlocal resources such as shared entanglement or quantum channels.

The initial quantum data-hiding scheme was shown through two-qubit orthogonal entangled states\cite{terh2001}.
In this scheme, one classical bit is asymptotically hidden under LOCC measurements but can be perfectly recovered once the players are allowed to perform global measurements.
Since then, a variety of quantum data-hiding schemes with different structural features and resource requirements have been developed\cite{lupo2016,lami2021,ha20241,ha20251}.
Nonetheless, it remains an open problem which types of quantum correlations are necessary for quantum data hiding.

Surprisingly, quantum data hiding does not necessarily require entanglement\cite{egge2002,ha20252,mele2025}.
The first data-hiding scheme based on separable states employs nonorthogonal states, in which one classical bit is hidden and recovered through asymptotic procedures\cite{egge2002}.
More recently, it was also shown that a quantum data-hiding scheme can be constructed using orthogonal separable states\cite{ha20252,mele2025}.
However, it remains unclear whether quantum data hiding can be achieved using two-qubit orthogonal separable states, that is, orthogonal separable states of the lowest possible dimension.

Here, we present a quantum data-hiding scheme based on two-qubit orthogonal separable states exhibiting local indistinguishability.
In our scheme, the local indistinguishability of the states is exploited to asymptotically hide one classical bit under LOCC constraints by repeatedly employing separable states of the lowest possible dimension.
As the number of repetitions increases, the information accessible about the hidden bit using LOCC measurements decays exponentially, approaching the level of random guessing.
This feature makes the practical realization of our scheme significantly more attainable.
Moreover, the orthogonality of the states guarantees the global distinguishability to perfectly recover the hidden bit even without invoking any asymptotic procedure when full cooperation is allowed and global measurements are permitted.

This paper is organized as follows.
In Sec.~\ref{ssec:qsd}, we recall the definitions and properties of two-party quantum state discrimination under LOCC measurements.
In Sec.~\ref{ssec:qdh}, we introduce a two-party data-hiding scheme using two orthogonal quantum states.
In Sec.~\ref{sec:sdh}, we present a sufficient condition under which a two-party quantum state ensemble can be used to construct a scheme that hides one classical bit. 
We illustrate this condition with examples of two-qubit orthogonal separable state ensembles. 
In Sec.~\ref{sec:disc}, we conclude by summarizing our findings and outlining possible directions for further research.

\section{Preliminaries}\label{sec:pre}
A two-party quantum state is expressed by a positive-semidefinite operator $\rho\succeq0$ with unit trace $\Tr\rho=1$, acting on a two-party Hilbert space $\mathcal{H}=\mathbb{C}^{d_{\mathrm{A}}}\otimes\mathbb{C}^{d_{\mathrm{B}}}$ with integers $d_{\mathrm{A}},d_{\mathrm{B}}\geqslant2$.
A measurement is described by a set of positive-semidefinite operators $M_{i}\succeq0$ acting on $\mathcal{H}$ and satisfying $\sum_{i}M_{i}=\mathbbm{1}$, where $\mathbbm{1}$ is the identity operator acting on $\mathcal{H}$.
For a quantum state $\rho$, performing a measurement $\{M_{i}\}_{i}$ results in the outcome corresponding to $M_{i}$ with probability $\Tr(\rho M_{i})$.

\subsection{Two-party quantum state discrimination}\label{ssec:qsd}
Let us consider the situation of discriminating two-party quantum states $\rho_{0}$ and $\rho_{1}$ which are prepared with probabilities $\eta_{0}$ and $\eta_{1}$, respectively.
We represent this situation as an ensemble
\begin{equation}\label{eq:enb}
\mathcal{E}=\{\eta_{0},\rho_{0};\eta_{1},\rho_{1}\}.
\end{equation}

To guess the state prepared from the ensemble $\mathcal{E}$ in Eq.~\eqref{eq:enb}, we employ the decision rule using a measurement
\begin{equation}\label{eq:mst}
\mathcal{M}=\{M_{0},M_{1}\},
\end{equation}
where obtaining the measurement outcomes corresponding to $M_{0}$ or $M_{1}$ indicates that the prepared state is inferred to be $\rho_{0}$ or $\rho_{1}$, respectively.
The \emph{minimum-error discrimination} of $\mathcal{E}$ is to achieve the maximum average probability of correctly guessing the prepared state from $\mathcal{E}$, that is,
\begin{equation}\label{eq:prgm}
p_{\mathsf{G}}(\mathcal{E})=\max_{\mathcal{M}}\sum_{i\in\{0,1\}}\eta_{i}\Tr(\rho_{i}M_{i}),
\end{equation}
where the maximum is taken over all possible measurements\cite{hels1969}.

\begin{definition}\label{def:ppt}
A Hermitian operator $E$ on $\mathcal{H}$ is called \emph{positive partial transpose}(PPT) if the partially transposed operator is positive semidefinite, that is,
\begin{equation}\label{eq:ppt}
E^{\PT}\succeq0,
\end{equation}
where the superscript $\PT$ denotes the partial transposition\cite{pere1996,pptp}.
\end{definition}

A measurement is called an \emph{LOCC measurement} if it can be implemented by LOCC, and a measurement $\{M_{i}\}_{i}$ is called a \emph{PPT measurement} if $M_{i}$ is PPT for all $i$.
It is known that any LOCC measurement is a PPT measurement\cite{chit2014}.
Under the restriction to LOCC measurements, we denote by $p_{\mathsf{L}}(\mathcal{E})$ the maximum average probability of correctly guessing the prepared state from $\mathcal{E}$ in Eq.~\eqref{eq:enb}, that is,
\begin{equation}\label{eq:ploc}
p_{\mathsf{L}}(\mathcal{E})=\max_{\mathsf{LOCC}\,\mathcal{M}}\sum_{i\in\{0,1\}}\eta_{i}\Tr(\rho_{i}M_{i}).
\end{equation}
Similarly, we denote
\begin{equation}\label{eq:prpt}
p_{\mathsf{PPT}}(\mathcal{E})=\max_{\mathsf{PPT}\,\mathcal{M}}\sum_{i\in\{0,1\}}\eta_{i}\Tr(\rho_{i}M_{i}),
\end{equation}
where the maximum is taken over all possible PPT measurements.
From the definitions, we have
\begin{equation}\label{eq:inpm}
\tfrac{1}{2}\leqslant\max\{\eta_{0},\eta_{1}\}\leqslant p_{\mathsf{L}}(\mathcal{E})
\leqslant p_{\mathsf{PPT}}(\mathcal{E})
\leqslant p_{\mathsf{G}}(\mathcal{E}),
\end{equation}
where the second inequality is due to the fact that simply guessing the state with the largest probability is an LOCC measurement.

The following proposition provides a method to obtain an upper bound of $p_{\mathsf{PPT}}(\mathcal{E})$ in Eq.~\eqref{eq:prpt}\cite{ha20252}.
\begin{proposition}\label{pro:upbd}
For a two-party quantum state ensemble $\mathcal{E}=\{\eta_{0},\rho_{0};\eta_{1},\rho_{1}\}$ satisfying
\begin{equation}\label{eq:pinv}
\Lambda_{\mathcal{E}}^{\PT}=\Lambda_{\mathcal{E}}
\end{equation}
where
\begin{equation}\label{eq:lame}
\Lambda_{\mathcal{E}}=\eta_{0}\rho_{0}-\eta_{1}\rho_{1},
\end{equation}
we have
\begin{equation}\label{eq:minq}
p_{\mathsf{PPT}}(\mathcal{E})=\tfrac{1}{2}+\min\Tr|H|,
\end{equation}
where $\Tr|\cdot|$ denotes the trace norm and the minimum is taken over all possible Hermitian operators $H$ with
\begin{equation}\label{eq:hhpt}
H+H^{\PT}=\Lambda_{\mathcal{E}}.
\end{equation}
\end{proposition}

For the ensemble $\mathcal{E}$ in Eq.~\eqref{eq:enb}, Proposition~\ref{pro:upbd} tells us that $\tfrac{1}{2}+\Tr|H|$ is an upper bound of $p_{\mathsf{PPT}}(\mathcal{E})$ for any Hermitian operator $H$ with Eq.~\eqref{eq:hhpt}.
We note that Eq.~\eqref{eq:pinv} is equivalent to Eq.~\eqref{eq:hhpt}; that is, Eq.~\eqref{eq:pinv} holds if and only if there exists a Hermitian operator $H$ satisfying Eq.~\eqref{eq:hhpt}.

\subsection{Two-party quantum data hiding}\label{ssec:qdh}
In this subsection, we first introduce the notion of a \emph{multifold ensemble} and then present a quantum data-hiding scheme that conceals one classical bit between two players, Alice and Bob.
In this scheme, the hidden bit can be perfectly recovered using an appropriate global measurement, whereas any LOCC measurement reveals only an arbitrarily small amount of information about it.

For the ensemble $\mathcal{E}$ in Eq.~\eqref{eq:enb} and a positive integer $L$, we define
\begin{equation}\label{eq:ebrb}
\eta_{\vec{b}}=\eta_{b_{1}}\times\cdots\times\eta_{b_{L}},~
\rho_{\vec{b}}=\rho_{b_{1}}\otimes\cdots\otimes\rho_{b_{L}},~
\vec{b}=(b_{1},\ldots,b_{L})\in\mathbb{Z}_{2}^{L}
\end{equation}
where $\mathbb{Z}_{2}^{L}$ is the Cartesian product of $L$ copies of $\{0,1\}$, that is,
\begin{equation}\label{eq:sztl}
\mathbb{Z}_{2}^{L}=\{(b_{1},\ldots,b_{L})\mid b_{1},\ldots,b_{L}\in\{0,1\}\}.
\end{equation}
We then write $\mathcal{E}^{\otimes L}$ for the \emph{$L$-fold ensemble} of $\mathcal{E}$, that is,
\begin{equation}\label{eq:lfeb}
\mathcal{E}^{\otimes L}=\{\eta_{\vec{b}},\rho_{\vec{b}}\}_{\vec{b}\in\mathbb{Z}_{2}^{L}},
\end{equation}
where each state $\rho_{\vec{b}}$ is prepared with probability $\eta_{\vec{b}}$, for $\vec{b}\in\mathbb{Z}_{2}^{L}$.
For each vector $\vec{b}=(b_{1},\ldots,b_{L})\in\mathbb{Z}_{2}^{L}$, we define $\omega_{2}(\vec{b})$ as the modulo-2 sum of all the entries in $\vec{b}$, that is,
\begin{equation}\label{eq:mtsb}
\omega_{2}(\vec{b})=b_{1}\oplus\cdots\oplus b_{L}\in\{0,1\}
\end{equation}
where $\oplus$ is the modulo-2 addition.

Let us consider the situation of guessing $\omega_{2}(\vec{b})$ of the prepared state $\rho_{\vec{b}}$ from the ensemble $\mathcal{E}^{\otimes L}$ in Eq.~\eqref{eq:lfeb}.
This situation is equivalent to guessing the prepared state from the two-state ensemble 
\begin{equation}\label{eq:cgeb}
\mathcal{E}^{(L)}=\{\eta_{0}^{(L)},\rho_{0}^{(L)};\eta_{1}^{(L)},\rho_{1}^{(L)}\}
\end{equation}
where the states $\rho_{0}^{(L)}$ and $\rho_{1}^{(L)}$ are prepared with the probabilities $\eta_{0}^{(L)}$ and $\eta_{1}^{(L)}$, respectively, and
\begin{equation}\label{eq:cgsp}
\eta_{i}^{(L)}=\sum_{\substack{\vec{b}\in\mathbb{Z}_{2}^{L}\\ \omega_{2}(\vec{b})=i}}\eta_{\vec{b}},~
\rho_{i}^{(L)}=\frac{1}{\eta_{i}^{(L)}}\sum_{\substack{\vec{b}\in\mathbb{Z}_{2}^{L}\\ \omega_{2}(\vec{b})=i}}\eta_{\vec{b}}\rho_{\vec{b}},~
i=0,1.
\end{equation}
Note that
\begin{equation}\label{eq:lael}
\Lambda_{\mathcal{E}^{(L)}}=\eta_{0}^{(L)}\rho_{0}^{(L)}-\eta_{1}^{(L)}\rho_{1}^{(L)}=(\eta_{0}\rho_{0}-\eta_{1}\rho_{1})^{\otimes L}=\Lambda_{\mathcal{E}}^{\otimes L},
\end{equation}
where the first and last equalities follow from the definition of $\Lambda_{\mathcal{E}}$ in Eq.~\eqref{eq:lame} and the second equality holds because Eq.~\eqref{eq:cgsp} implies
\begin{align}\label{eq:eril}
\eta_{i}^{(L)}\rho_{i}^{(L)}=\frac{1}{2}\Big[&(\eta_{0}\rho_{0}+\eta_{1}\rho_{1})^{\otimes L}
+(-1)^{i}(\eta_{0}\rho_{0}-\eta_{1}\rho_{1})^{\otimes L}\Big],~i=0,1.
\end{align}

For the ensemble $\mathcal{E}$ in Eq.~\eqref{eq:enb}, the following proposition provides a sufficient condition for $p_{\mathsf{PPT}}(\mathcal{E}^{(L)})$ in Eq.~\eqref{eq:prpt} to be monotonically decreasing as $L$ increases\cite{ha20252}.
\begin{proposition}\label{pro:mode}
For an ensemble $\mathcal{E}=\{\eta_{0},\rho_{0};\eta_{1},\rho_{1}\}$ with Eq.~\eqref{eq:pinv}, $p_{\mathsf{PPT}}(\mathcal{E}^{(L)})$ is monotonically decreasing as $L$ increases.
\end{proposition}

Now, we consider an \emph{orthogonal} quantum state ensemble $\mathcal{E}=\{\eta_{0},\rho_{0};\eta_{1},\rho_{1}\}$ satisfying
\begin{equation}\label{eq:plhf}
\lim_{L\rightarrow\infty}p_{\mathsf{PPT}}(\mathcal{E}^{(L)})=\tfrac{1}{2}.
\end{equation}
By using the ensemble $\mathcal{E}$, we can construct a quantum data-hiding scheme that conceals one classical bit\cite{ha20241}.
The scheme is given as follows.

The hider, acting as the dealer, first prepares a state $\rho_{b_{1}}$ from $\mathcal{E}$ with the corresponding probability $\eta_{b_{1}}$ for $b_{1}\in\{0,1\}$ and distributes it between Alice and Bob. 
The hider repeats this process $L$ times so that a state $\rho_{b_{l}}$ is prepared with probability $\eta_{b_{l}}$ in the $l$th repetition for $l=1,\ldots,L$.
This whole procedure is equivalent to the situation where the hider first prepares a state $\rho_{\vec{b}}$ from the $L$-fold ensemble of $\mathcal{E}$ with the corresponding probability $\eta_{\vec{b}}$ for $\vec{b}=(b_{1},\ldots,b_{L})\in\mathbb{Z}_{2}^{L}$ and distributes it between Alice and Bob. 
For the state $\rho_{\vec{b}}$ shared between Alice and Bob, the classical bit $\omega_{2}(\vec{b})$ is denoted by $y$, that is,
\begin{equation}\label{eq:clby}
y=\omega_{2}(\vec{b}).
\end{equation}
To conceal one classical bit $x\in\{0,1\}$, the hider broadcasts to Alice and Bob the classical information $z$ where
\begin{equation}\label{eq:clbz}
z=x\oplus y.
\end{equation}

As the bit $z$ in Eq.~\eqref{eq:clbz} is broadcast to Alice and Bob, guessing the bit $x$ becomes equivalent to guessing the bit $y$ in Eq.~\eqref{eq:clby}.
Since guessing $y$ is equivalent to guessing the prepared state from $\mathcal{E}^{(L)}$ in Eq.~\eqref{eq:cgeb}, the maximum average probability of correctly guessing $x$ is equal to that of correctly guessing the prepared state from $\mathcal{E}^{(L)}$, that is, $p_{\mathsf{G}}(\mathcal{E}^{(L)})$.
Similarly, the maximum average probability of correctly guessing $x$ using only LOCC measurements becomes $p_{\mathsf{L}}(\mathcal{E}^{(L)})$.

Since the orthogonality of the states in $\mathcal{E}$ implies mutual orthogonality among the states of $\mathcal{E}^{\otimes L}$, it follows from the definitions of $\rho_{0}^{(L)}$ and $\rho_{1}^{(L)}$ in Eq.~\eqref{eq:cgsp} that the states of $\mathcal{E}^{(L)}$ are orthogonal, and therefore
\begin{equation}\label{eq:pgoe}
p_{\mathsf{G}}(\mathcal{E}^{(L)})=1.
\end{equation}
In other words, the classical bit $x$ is fully revealed when Alice and Bob are allowed to collaborate and perform global measurements.
On the other hand, Eq.~\eqref{eq:plhf} implies
\begin{equation}\label{eq:pthf}
\lim_{L\rightarrow\infty}p_{\mathsf{L}}(\mathcal{E}^{(L)})=\tfrac{1}{2},
\end{equation}
since Inequality~\eqref{eq:inpm} shows that $p_{\mathsf{L}}(\mathcal{E}^{(L)})$ is bounded below by $1/2$ and above by $p_{\mathsf{PPT}}(\mathcal{E}^{(L)})$ for each positive integer $L$.
Thus, $p_{\mathsf{L}}(\mathcal{E}^{(L)})$ can be arbitrarily close to $1/2$ by choosing $L$ sufficiently large.
In other words, the globally accessible bit $x$ is perfectly concealed asymptotically if Alice and Bob are only able to use LOCC measurements.

\section{Quantum data-hiding scheme using two-qubit separable state ensemble}\label{sec:sdh}
In this section, we present \emph{two-qubit} separable state ensembles that can be used to construct a quantum data-hiding scheme.
To this end, we first provide a sufficient condition that guarantees the convergence~\eqref{eq:plhf}.

For the ensemble $\mathcal{E}$ in Eq.~\eqref{eq:enb}, the following theorem establishes a sufficient condition under which $p_{\mathsf{PPT}}(\mathcal{E}^{(L)})$ converges exponentially to $1/2$ as $L$ increases.
\begin{theorem}\label{thm:cppt}
For an ensemble $\mathcal{E}=\{\eta_{0},\rho_{0};\eta_{1},\rho_{1}\}$ with Eq.~\eqref{eq:pinv}, if there exist a positive integer $k$ and a Hermitian operator $H$ satisfying 
\begin{subequations}\label{eq:acek}
\begin{align}
H+H^{\PT}=\Lambda_{\mathcal{E}}^{\otimes k},\label{eq:hlek}\\
4\Tr|H|\Tr|H^{\PT}|<1,\label{eq:fhho}
\end{align}
\end{subequations}
then
\begin{equation}\label{eq:mupb}
p_{\mathsf{PPT}}(\mathcal{E}^{(L)})\leqslant\tfrac{1}{2}+\tfrac{1}{2}\big(4\Tr|H|\Tr|H^{\PT}|\big)^{\frac{L-k+1}{2k}}
\end{equation}
for any positive integer $L$.
In this case, $p_{\mathsf{PPT}}(\mathcal{E}^{(L)})$ converges exponentially to $1/2$ as $L$ increases.
\end{theorem}

To prove Theorem~\ref{thm:cppt}, we use the following lemma to obtain an upper bound on $p_{\mathsf{PPT}}(\mathcal{E}^{(mk)})$ for a fixed positive integer $k$ and any positive integer $m$.
The proof of Lemma~\ref{lem:scpt} is given in Appendix~\ref{app:scpt}.
\begin{lemma}\label{lem:scpt}
For an ensemble $\mathcal{E}=\{\eta_{0},\rho_{0};\eta_{1},\rho_{1}\}$, if there exist a positive integer $k$ and a Hermitian operator $H$ with Eq.~\eqref{eq:hlek}, then 
\begin{equation}\label{eq:upkr}
p_{\mathsf{PPT}}(\mathcal{E}^{(mk)})\leqslant
\tfrac{1}{2}+\tfrac{1}{2}\big(4\Tr|H|\Tr|H^{\PT}|\big)^{\frac{m}{2}}
\end{equation}
for any positive integer $m$.
\end{lemma}

\begin{proof}[Proof of Theorem~\ref{thm:cppt}]
For any positive integer $L<k$, Inequality~\eqref{eq:mupb} holds because
\begin{equation}\label{eq:pulk}
p_{\mathsf{PPT}}(\mathcal{E}^{(L)})\leqslant 1\leqslant
\tfrac{1}{2}+\tfrac{1}{2}\big(4\Tr|H|\Tr|H^{\PT}|\big)^{\frac{L-k+1}{2k}},
\end{equation}
where the second inequality is from Inequality~\eqref{eq:fhho} and $-\frac{1}{2}<\frac{L-k+1}{2k}\leqslant 0$.
For any positive integer $L\geqslant k$, Inequality~\eqref{eq:mupb} is also satisfied because
\begin{align}\label{eq:mptb}
p_{\mathsf{PPT}}(\mathcal{E}^{(L)})
&\leqslant p_{\mathsf{PPT}}(\mathcal{E}^{(\lfloor\frac{L}{k}\rfloor k)})\nonumber\\
&\leqslant\tfrac{1}{2}+\tfrac{1}{2}\big(4\Tr|H|\Tr|H^{\PT}|\big)^{\frac{1}{2}\lfloor\frac{L}{k}\rfloor}\nonumber\\
&\leqslant\tfrac{1}{2}+\tfrac{1}{2}\big(4\Tr|H|\Tr|H^{\PT}|\big)^{\frac{1}{2}(\frac{L}{k}-\frac{k-1}{k})}\nonumber\\
&=\tfrac{1}{2}+\tfrac{1}{2}\big(4\Tr|H|\Tr|H^{\PT}|\big)^{\frac{L-k+1}{2k}},
\end{align}
where $\lfloor\cdot\rfloor$ is the floor function, the first inequality is from Proposition~\ref{pro:mode} and Eq.~\eqref{eq:pinv} together with $L\geqslant \lfloor\frac{L}{k}\rfloor k$, the second inequality is from Lemma~\ref{lem:scpt} and Eq.~\eqref{eq:hlek}, and the last inequality is from Inequality~\eqref{eq:fhho} and $\lfloor\frac{L}{k}\rfloor\geqslant\frac{L}{k}-\frac{k-1}{k}$.
Moreover, it follows from Inequalities~\eqref{eq:inpm}, \eqref{eq:fhho} and \eqref{eq:mupb} that $p_{\mathsf{PPT}}(\mathcal{E}^{(L)})$ converges exponentially to $1/2$ as $L$ increases.
\qedhere
\end{proof}

\begin{corollary}\label{cor:scpt}
For an ensemble $\mathcal{E}=\{\eta_{0},\rho_{0};\eta_{1},\rho_{1}\}$, if there exists a Hermitian operator $H$ satisfying Eq.~\eqref{eq:hhpt} and 
\begin{subequations}\label{eq:idsc}
\begin{align}
&\Tr|H|+\Tr|H^{\PT}|\leqslant 1,\label{eq:idsf}\\
&\Tr|H|<\tfrac{1}{2},\label{eq:idss}
\end{align}
\end{subequations}
then 
\begin{equation}\label{eq:smup}
p_{\mathsf{PPT}}(\mathcal{E}^{(L)})\leqslant\tfrac{1}{2}+\tfrac{1}{2}\big(4\Tr|H|\Tr|H^{\PT}|\big)^{\frac{L}{2}}
\end{equation}
for any positive integer $L$.
In this case, $p_{\mathsf{PPT}}(\mathcal{E}^{(L)})$ converges exponentially to $1/2$ as $L$ increases.
\end{corollary}
\begin{proof}
Equation~\eqref{eq:hhpt} immediately implies that Eq.~\eqref{eq:pinv} is satisfied and that Eq.~\eqref{eq:hlek} holds for $k=1$. 
Inequality~\eqref{eq:fhho} holds since
\begin{equation}\label{eq:incc}
4\Tr|H|\Tr|H^{\PT}|\leqslant 4\Tr|H|(1-\Tr|H|)
=1-(1-2\Tr|H|)^{2}<1,
\end{equation}
where the first and second inequalities are from Inequalities~\eqref{eq:idsf} and \eqref{eq:idss}, respectively.
Thus, Theorem~\ref{thm:cppt} leads us to Inequality~\eqref{eq:smup}.
Moreover, $p_{\mathsf{PPT}}(\mathcal{E}^{(L)})$ converges exponentially to $1/2$ as $L$ increases.
\qedhere
\end{proof}

Now, we provide the following examples of \emph{two-qubit} state ensembles consisting of two orthogonal separable states. 
These ensembles can be used to construct a quantum data-hiding scheme that conceals one classical bit.
\begin{example}\label{ex:tqbt}
Let us consider the $2\otimes 2$ orthogonal separable state ensemble $\mathcal{E}=\{\eta_{0},\rho_{0};\eta_{1},\rho_{1}\}$ consisting of 
\begin{align}\label{eq:mrex}
\eta_{0}=\tfrac{1}{2},~\rho_{0}=&\tfrac{1}{2}\ket{0}\!\bra{0}\otimes\ket{+_{\theta}}\!\bra{+_{\theta}}
+\tfrac{1}{2}\ket{+_{\theta}}\!\bra{+_{\theta}}\otimes\ket{0}\!\bra{0},\nonumber\\
\eta_{1}=\tfrac{1}{2},~\rho_{1}=&\tfrac{1}{2}\ket{1}\!\bra{1}\otimes\ket{1}\!\bra{1}
+\tfrac{1}{2}\ket{-_{\theta}}\!\bra{-_{\theta}}\otimes\ket{-_{\theta}}\!\bra{-_{\theta}}
\end{align}
where
\begin{align}\label{eq:pmsd}
\ket{+_{\theta}}&=\cos\theta\ket{0}+\sin\theta\ket{1},\nonumber\\
\ket{-_{\theta}}&=\sin\theta\ket{0}-\cos\theta\ket{1},~0\leqslant\theta\leqslant\tfrac{\pi}{3}.
\end{align}
Note that the states $\rho_{0}$ and $\rho_{1}$ are orthogonal due to $\braket{+_{\theta}|-_{\theta}}=0$.
We also note that Eq.~\eqref{eq:pinv} is satisfied by the states and probabilities in Eq.~\eqref{eq:mrex} because $\ket{\pm_{\theta}}\!\bra{\pm_{\theta}}$ is invariant under transposition\cite{pptp}.
\end{example}

By using the $2\otimes2$ orthonormal basis $\{\ket{\psi_{i}^{+}},\ket{\psi_{i}^{-}}\}_{i\in\{0,1\}}$,
\begin{align}\label{eq:psii}
\ket{\psi_{0}^{+}}&=\tfrac{1}{\sqrt{2}}(\ket{00}+\ket{11}),\nonumber\\
\ket{\psi_{1}^{+}}&=\tfrac{1}{\sqrt{2}}(\ket{01}-\ket{10}),\nonumber\\
\ket{\psi_{0}^{-}}&=\tfrac{\cos\theta}{\sqrt{2}}(\ket{00}-\ket{11})
+\tfrac{\sin\theta}{\sqrt{2}}(\ket{01}+\ket{10}),\nonumber\\
\ket{\psi_{1}^{-}}&=\tfrac{\sin\theta}{\sqrt{2}}(\ket{00}-\ket{11})
-\tfrac{\cos\theta}{\sqrt{2}}(\ket{01}+\ket{10}),
\end{align}
we can define the $4\otimes4$ orthonormal basis $\{\ket{\Psi_{i}^{+}},\ket{\Psi_{i}^{-}}\}_{i=0}^{7}$,
\begin{align}\label{eq:cpsi}
&\ket{\Psi_{0}^{\pm}}=\tfrac{1}{\sqrt{2}}(\ket{\psi_{0}^{+}}\otimes\ket{\psi_{0}^{\pm}}
\mp\ket{\psi_{0}^{-}}\otimes\ket{\psi_{0}^{\mp}}),\nonumber\\
&\ket{\Psi_{1}^{\pm}}=\tfrac{1}{\sqrt{2}}(\ket{\psi_{0}^{+}}\otimes\ket{\psi_{0}^{\pm}}
\pm\ket{\psi_{0}^{-}}\otimes\ket{\psi_{0}^{\mp}}),\nonumber\\
&\ket{\Psi_{2}^{\pm}}=\ket{\psi_{1}^{+}}\otimes\ket{\psi_{1}^{\pm}},\nonumber\\
&\ket{\Psi_{3}^{\pm}}=\ket{\psi_{1}^{-}}\otimes\ket{\psi_{1}^{\mp}},\nonumber\\
&\ket{\Psi_{4}^{\pm}}=\big(\tfrac{\cos\theta}{\sqrt{1+\cos^{2}\theta}}\ket{\psi_{0}^{\pm}}\pm\tfrac{1}{\sqrt{1+\cos^{2}\theta}}\ket{\psi_{0}^{\mp}}\big)\otimes\ket{\psi_{1}^{+}},\nonumber\\
&\ket{\Psi_{5}^{\pm}}=\big(\tfrac{\cos\theta}{\sqrt{1+\cos^{2}\theta}}\ket{\psi_{0}^{\mp}}\mp\tfrac{1}{\sqrt{1+\cos^{2}\theta}}\ket{\psi_{0}^{\pm}}\big)\otimes\ket{\psi_{1}^{-}},\nonumber\\
&\ket{\Psi_{6}^{\pm}}=\ket{\psi_{1}^{+}}\otimes\big(\tfrac{\cos\theta}{\sqrt{1+\cos^{2}\theta}}\ket{\psi_{0}^{\pm}}\pm\tfrac{1}{\sqrt{1+\cos^{2}\theta}}\ket{\psi_{0}^{\mp}}\big),\nonumber\\
&\ket{\Psi_{7}^{\pm}}=\ket{\psi_{1}^{-}}\otimes\big(\tfrac{\cos\theta}{\sqrt{1+\cos^{2}\theta}}\ket{\psi_{0}^{\mp}}\mp\tfrac{1}{\sqrt{1+\cos^{2}\theta}}\ket{\psi_{0}^{\pm}}\big).
\end{align}
By tedious but straightforward calculations, we can show that
\begin{align}\label{eq:ltwd}
\Lambda_{\mathcal{E}}^{\otimes 2}=&
\tfrac{\sin^{4}\theta-4\cos^{2}\theta}{16}(\ket{\Psi_{0}^{+}}\!\bra{\Psi_{0}^{+}}-\ket{\Psi_{0}^{-}}\!\bra{\Psi_{0}^{-}})\nonumber\\
&-\tfrac{\cos\theta\sin^{2}\theta}{4}(\ket{\Psi_{0}^{+}}\!\bra{\Psi_{0}^{-}}+\ket{\Psi_{0}^{-}}\!\bra{\Psi_{0}^{+}})\nonumber\\
&+\tfrac{(1+\cos^{2}\theta)^{2}}{16}(\ket{\Psi_{1}^{+}}\!\bra{\Psi_{1}^{+}}-\ket{\Psi_{1}^{-}}\!\bra{\Psi_{1}^{-}})\nonumber\\
&+\tfrac{\sin^{4}\theta}{16}{\textstyle\sum_{i=2}^{3}}(\ket{\Psi_{i}^{+}}\!\bra{\Psi_{i}^{+}}-\ket{\Psi_{i}^{-}}\!\bra{\Psi_{i}^{-}})\nonumber\\
&+\tfrac{1-\cos^{4}\theta}{16}{\textstyle\sum_{i=4}^{7}}(\ket{\Psi_{i}^{+}}\!\bra{\Psi_{i}^{+}}-\ket{\Psi_{i}^{-}}\!\bra{\Psi_{i}^{-}}).
\end{align}

For $k=2$ and the Hermitian operator
\begin{align}\label{eq:expd}
H=&\tfrac{\sin^{4}\theta-2\cos^{2}\theta}{16}
(\ket{\Psi_{0}^{+}}\!\bra{\Psi_{0}^{+}}-\ket{\Psi_{0}^{-}}\!\bra{\Psi_{0}^{-}})\nonumber\\
&-\tfrac{\cos\theta\sin^{2}\theta}{8}
(\ket{\Psi_{0}^{+}}\!\bra{\Psi_{0}^{-}}+\ket{\Psi_{0}^{-}}\!\bra{\Psi_{0}^{+}})\nonumber\\
&+\tfrac{1+\cos^{4}\theta}{16}
(\ket{\Psi_{1}^{+}}\!\bra{\Psi_{1}^{+}}-\ket{\Psi_{1}^{-}}\!\bra{\Psi_{1}^{-}})\nonumber\\
&+\tfrac{1-\cos^{4}\theta}{32}{\textstyle\sum_{i=4}^{7}}(\ket{\Psi_{i}^{+}}\!\bra{\Psi_{i}^{+}}-\ket{\Psi_{i}^{-}}\!\bra{\Psi_{i}^{-}}),
\end{align}
it is straightforward to check Eq.~\eqref{eq:hlek} and its partial transposition
\begin{align}\label{eq:exhp}
H^{\PT}=&-\tfrac{\cos^{2}\theta}{8}(\ket{\Psi_{0}^{+}}\!\bra{\Psi_{0}^{+}}-\ket{\Psi_{0}^{-}}\!\bra{\Psi_{0}^{-}})\nonumber\\
&-\tfrac{\cos\theta\sin^{2}\theta}{8}(\ket{\Psi_{0}^{+}}\!\bra{\Psi_{0}^{-}}+\ket{\Psi_{0}^{-}}\!\bra{\Psi_{0}^{+}})\nonumber\\
&+\tfrac{\cos^{2}\theta}{8}(\ket{\Psi_{1}^{+}}\!\bra{\Psi_{1}^{+}}-\ket{\Psi_{1}^{-}}\!\bra{\Psi_{1}^{-}})\nonumber\\
&+\tfrac{\sin^{4}\theta}{16}{\textstyle\sum_{i=2}^{3}}(\ket{\Psi_{i}^{+}}\!\bra{\Psi_{i}^{+}}-\ket{\Psi_{i}^{-}}\!\bra{\Psi_{i}^{-}})\nonumber\\
&+\tfrac{1-\cos^{4}\theta}{32}{\textstyle\sum_{i=4}^{7}}(\ket{\Psi_{i}^{+}}\!\bra{\Psi_{i}^{+}}-\ket{\Psi_{i}^{-}}\!\bra{\Psi_{i}^{-}}).
\end{align}
From Eqs.~\eqref{eq:expd} and \eqref{eq:exhp}, we can verify that
\begin{align}\label{eq:metn}
\Tr|H|&=\tfrac{3-\cos^{4}\theta+\sqrt{4\cos^{4}\theta+\sin^{8}\theta}}{8}=:f_{0}(\theta),\nonumber\\
\Tr|H^{\PT}|&=\tfrac{1+\sin^{2}\theta+\cos\theta\sqrt{\cos^{2}\theta+\sin^{4}\theta}}{4}=:f_{1}(\theta).
\end{align}

\begin{figure}[!tt]
\centerline{\includegraphics*[bb=20 20 410 305,scale=0.62]{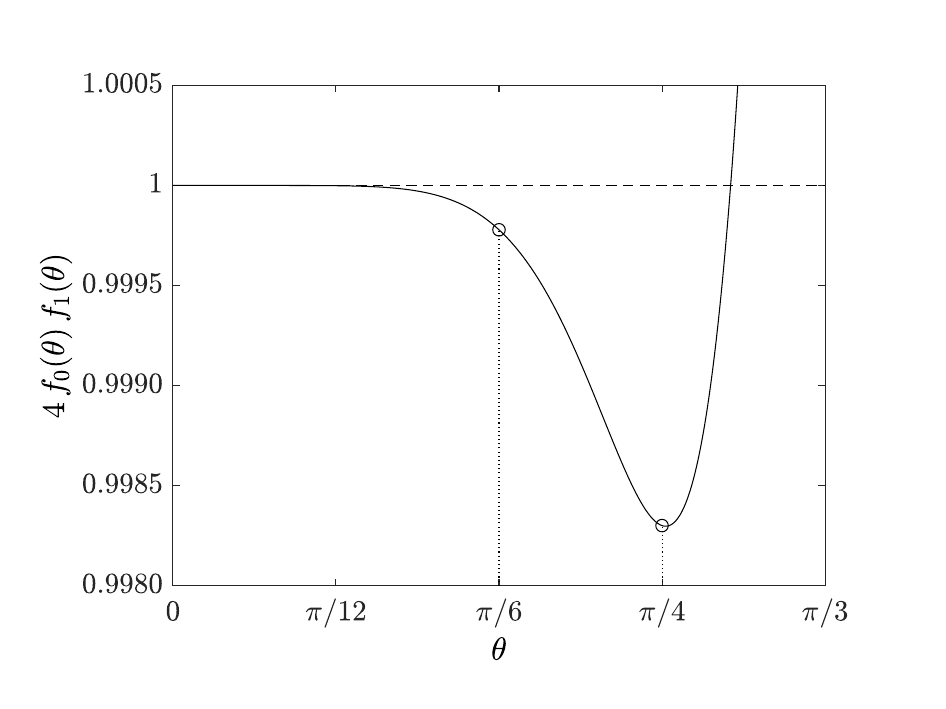}}
\caption{
The behavior of $4f_{0}(\theta)f_{1}(\theta)$ on the interval $0\leqslant\theta\leqslant\frac{\pi}{3}$. The function takes values approximately $0.9998$ and $0.9983$ at $\theta=\pi/6$ and $\theta=\pi/4$, respectively. When $4f_{0}(\theta)f_{1}(\theta)<1$, the two-qubit separable state ensemble $\mathcal{E}$ in Example~\ref{ex:tqbt} can be used to construct a one-bit hiding scheme. 
}\label{fig:fplt}
\end{figure}

For the case when
\begin{equation}\label{eq:lmcd}
4f_{0}(\theta)f_{1}(\theta)<1,
\end{equation}
it follows from Theorem~\ref{thm:cppt} that $p_{\mathsf{PPT}}(\mathcal{E}^{(L)})$ converges exponentially to $1/2$ as $L$ increases, that is,
\begin{equation}\label{eq:mube}
p_{\mathsf{PPT}}(\mathcal{E}^{(L)})\leqslant\tfrac{1}{2}+\tfrac{1}{2}\big[4f_{0}(\theta)f_{1}(\theta)\big]^{\frac{L-1}{4}}
\end{equation}
for any positive integer $L$.
In this case, the two-qubit separable state ensemble $\mathcal{E}$ in Example~\ref{ex:tqbt} can be used to construct a quantum data-hiding scheme that conceals one classical bit. 
The details of a quantum data-hiding scheme are given in Sec.~\ref{ssec:qdh}.
Figure~\ref{fig:fplt} illustrates the left-hand side $4f_{0}(\theta)f_{1}(\theta)$ of Inequality~\eqref{eq:lmcd} as a function of $\theta$.
In particular, we have
\begin{align}\label{eq:iptf}
4f_{0}(\theta)f_{1}(\theta)
=
\begin{dcases}
\tfrac{(39+\sqrt{577})(10+\sqrt{39})}{1024}\approx0.9998,&\theta=\tfrac{\pi}{6},\\
\tfrac{(11+\sqrt{17})(6+\sqrt{6})}{128}\approx0.9983,&\theta=\tfrac{\pi}{4}.
\end{dcases}
\end{align}


\section{Discussion}\label{sec:disc}
We have considered two-party quantum state discrimination and provided a quantum data-hiding scheme using two-qubit orthogonal separable states. 
We have first provided a bound on the optimal local discrimination for two-party state ensembles, and established a sufficient condition under which a two-party quantum state ensemble can be used to construct a data-hiding scheme. 
We have illustrated this condition through examples of two-qubit state ensembles consisting of two orthogonal separable states.

In our data-hiding scheme, one classical bit can be asymptotically hidden perfectly under LOCC constraints by repeatedly employing separable states of the lowest possible dimension. 
As the number of repetitions increases, the information accessible about the hidden bit under LOCC constraints decays exponentially, eventually approaching the level of random guessing. 
This feature makes the practical realization of our scheme significantly more attainable.
Moreover, full cooperation enabling global measurements allows the hidden bit to be perfectly recovered without any asymptotic method.

As our results provide one-bit hiding schemes based on two-qubit separable states, it is natural to extend our investigation to multi-bit hiding schemes constructed from multi-qubit separable states.
More broadly, it remains an open question whether every orthogonal quantum state ensemble exhibiting local indistinguishability can be used to construct a quantum data-hiding scheme.

\section*{Acknowledgments}
This work was supported by Korea Research Institute for defense Technology planning and advancement (KRIT) grant funded by Defense Acquisition Program Administration(DAPA)(KRIT-CT-23–031), a National Research Foundation of Korea(NRF) grant funded by the Korean government(Ministry of Science and ICT) (No.NRF2023R1A2C1007039), and the Institute for Information \& Communications Technology Planning \& Evaluation(IITP) grant funded by the Korean government(MSIP)(Grant No. RS-2025-02304540). JSK was supported by Creation of the Quantum Information Science R\&D Ecosystem(Grant No. 2022M3H3A106307411) through the National Research Foundation of Korea(NRF) funded by the Korean government(Ministry of Science and ICT).

\appendix
\appsection{Proof of Lemma~\ref{lem:scpt}}\label{app:scpt}
We define $H'$ as follows:
\begin{equation}\label{eq:hadf}
H'=
\begin{dcases}
H,&\Tr|H|\leqslant\Tr|H^{\PT}|,\\
H^{\PT},&\Tr|H|>\Tr|H^{\PT}|.
\end{dcases}
\end{equation}
From this definition and Eq.~\eqref{eq:hlek}, we have
\begin{equation}\label{eq:hplc}
H'+H'^{\PT}=H+H^{\PT}=\Lambda_{\mathcal{E}}^{\otimes k}.
\end{equation}

For a positive integer $m$, let us consider the Hermitian operator
\begin{equation}\label{eq:dfhp}
\tilde{H}=
\begin{dcases}
\sum_{r=0}^{(m-1)/2}\tilde{H}_{r},&m:\mathrm{odd},\\
\sum_{r=0}^{(m-2)/2}\tilde{H}_{r}+\frac{1}{2}\tilde{H}_{m/2},&m:\mathrm{even},
\end{dcases}
\end{equation}
where each $\tilde{H}_{r}$ is the sum of all $m$-fold tensor products of $H'$ and $H'^{\PT}$ with exactly $r$ factors equal to $H'^{\PT}$, that is,
\begin{equation}\label{eq:dfhr}
\tilde{H}_{r}=\sum_{\substack{(b_{1},\ldots,b_{m})\in\mathbb{Z}_{2}^{m}\\ b_{1}+\cdots+b_{m}=r}}\bigotimes_{l=1}^{m}\Big[(1-b_{l})H'+b_{l}H'^{\PT}\Big],
\end{equation}
for $r=0,1,\ldots,m$.
We prove Inequality~\eqref{eq:upkr} by showing that
\begin{subequations}\label{eq:setc}
\begin{align}
&p_{\mathsf{PPT}}(\mathcal{E}^{(mk)})\leqslant\tfrac{1}{2}+\Tr|\tilde{H}|,\label{eq:sefc}\\
&\Tr|\tilde{H}|\leqslant\tfrac{1}{2}\big(4\Tr|H|\Tr|H^{\PT}|\big)^{\frac{m}{2}}.\label{eq:sesc}
\end{align}
\end{subequations}

\begin{proof}[Proof of Inequality~\eqref{eq:sefc}]
For each $r=0,1,\ldots,m$, we can see from the definition of $\tilde{H}_{r}$ in Eq.~\eqref{eq:dfhr} that
\begin{equation}\label{eq:hrpt}
\tilde{H}_{r}^{\PT}=\tilde{H}_{m-r}.
\end{equation}
The Hermitian operator $\tilde{H}$ in Eq.~\eqref{eq:dfhp} satisfies
\begin{equation}\label{eq:hhpc}
\tilde{H}+\tilde{H}^{\PT}=\sum_{r=0}^{m}\tilde{H}_{r}
=(H'+H'^{\PT})^{\otimes m}
=(\Lambda_{\mathcal{E}}^{\otimes k})^{\otimes m}=\Lambda_{\mathcal{E}}^{\otimes mk}=\Lambda_{\mathcal{E}^{\otimes mk}},
\end{equation}
where the first equality is from Eq.~\eqref{eq:hrpt}, the second equality is from the definition of $\tilde{H}_{r}$ in Eq.~\eqref{eq:dfhr}, the third equality is from Eq.~\eqref{eq:hplc}, and the last equality is from Eq.~\eqref{eq:lael}.
Thus, Proposition~\ref{pro:upbd} leads us to Inequality~\eqref{eq:sefc}.
\qedhere
\end{proof}

\begin{proof}[Proof of Inequality~\eqref{eq:sesc}]
For each integer $r$ with $0\leqslant r\leqslant m/2$, we have
\begin{align}\label{eq:hrtn}
\Tr|\tilde{H}_{r}|&\leqslant\sum_{\substack{(b_{1},\ldots,b_{m})\in\mathbb{Z}_{2}^{m}\\ b_{1}+\cdots+b_{m}=r}}\prod_{l=1}^{m}\Tr\big|(1-b_{l})H'+b_{l}H'^{\PT}\big|\nonumber\\
&=\binom{m}{r}\big(\Tr|H'|\big)^{m-r}\big(\Tr|H'^{\PT}|\big)^{r}\nonumber\\
&\leqslant\binom{m}{r}\big(\Tr|H'|\big)^{\frac{m}{2}}\big(\Tr|H'^{\PT}|\big)^{\frac{m}{2}},
\end{align}
where $\binom{m}{r}$ denotes the binomial coefficient $\frac{m!}{r!(m-r)!}$, the first inequality is from the definition of $\tilde{H}_{r}$ in Eq.~\eqref{eq:dfhr} together with the triangle inequality of trace norm and $\Tr|A\otimes B|=\Tr|A|\Tr|B|$ for any Hermitian operators $A$ and $B$, and the second inequality is from $0\leqslant r\leqslant m/2$ and the definition of $H'$ in Eq.~\eqref{eq:hadf}.
For odd $m$, we have
\begin{align}\label{eq:cotn}
\Tr|\tilde{H}|&\leqslant\sum_{r=0}^{(m-1)/2}\Tr|\tilde{H}_{r}|\nonumber\\
&\leqslant\sum_{r=0}^{(m-1)/2}\binom{m}{r}\big(\Tr|H'|\big)^{\frac{m}{2}}\big(\Tr|H'^{\PT}|\big)^{\frac{m}{2}}\nonumber\\
&=2^{m-1}\big(\Tr|H'|\big)^{\frac{m}{2}}\big(\Tr|H'^{\PT}|\big)^{\frac{m}{2}}\nonumber\\
&=\tfrac{1}{2}\big(4\Tr|H'|\Tr|H'^{\PT}|\big)^{\frac{m}{2}}\nonumber\\
&=\tfrac{1}{2}\big(4\Tr|H|\Tr|H^{\PT}|\big)^{\frac{m}{2}},
\end{align}
where the first inequality is from the definition of $\tilde{H}$ in Eq.~\eqref{eq:dfhp} together with the triangle inequality of trace norm, the second inequality is from Inequality~\eqref{eq:hrtn}, the first equality is from $\sum_{r=0}^{m}\binom{m}{r}=2^{m}$ together with $\binom{m}{r}=\binom{m}{m-r}$ for each $r=0,1,\ldots,m$, and the last equality is from the definition of $H'$ in Eq.~\eqref{eq:hadf}.
For even $m$, we also have
\begin{align}\label{eq:cetn}
\Tr|\tilde{H}|\leqslant&\sum_{r=0}^{(m-2)/2}\Tr|\tilde{H}_{r}|+\frac{1}{2}\Tr|\tilde{H}_{\frac{m}{2}}|\nonumber\\
\leqslant&\Bigg[\sum_{r=0}^{(m-2)/2}\binom{m}{r}+\frac{1}{2}\binom{m}{\frac{m}{2}}\Bigg]
\big(\Tr|H'|\big)^{\frac{m}{2}}\big(\Tr|H'^{\PT}|\big)^{\frac{m}{2}}\nonumber\\
=&2^{m-1}\big(\Tr|H'|\big)^{\frac{m}{2}}\big(\Tr|H'^{\PT}|\big)^{\frac{m}{2}}\nonumber\\
=&\tfrac{1}{2}\big(4\Tr|H'|\Tr|H'^{\PT}|\big)^{\frac{m}{2}}\nonumber\\
=&\tfrac{1}{2}\big(4\Tr|H|\Tr|H^{\PT}|\big)^{\frac{m}{2}},
\end{align}
where the inequalities and equalities follow in a manner analogous to that of Eq.~\eqref{eq:cotn}.
Thus, Inequality~\eqref{eq:sesc} is satisfied.
\qedhere
\end{proof}

 

\begin{thebibliography}{99}
\bibitem{pptp}
The eigenvalues of a partially transposed operator do not depend on the choice of basis or the subsystem to be transposed. 
For simplicity, we consider the standard basis and the second system throughout this paper.
\end{thebibliography}
\end{document}